\documentstyle[12pt,a4,epsfig]{article}

\begin{document}
\section*{Sunyaev Zel`dovich effect studies with
MASTER}
\par~
\par\noindent\centerline{A.Tartari$^{1}$, G.Boella$^{1}$, M.Candotti$^{3}$, M.Gervasi$^{1}$,}
\par\noindent\centerline{V.Natale$^{3)}$, A.Passerini$^{1}$, G.Sironi$^{1}$, M.Zannoni$^{2}$}
\par~\par\centerline{$~^{(1)}$Dipartimento di Fisica G.Occhialini, Universit\'a degli
Studi}
\par\noindent\centerline{di Milano Bicocca, piazza della Scienza 3, 20126 Milano,
Italy}
\par\noindent\centerline{$~^{(2)}$I.A.S.F.-C.N.R., Sezione di Milano}
\par\noindent\centerline{via Bassini 15 , 20133 Milano, Italy}
\par\noindent\centerline{$~^{(3)}$I.R.A.-C.N.R., sezione di Firenze}
\par\noindent\centerline{ largo Enrico Fermi 5,50125 Firenze, Italy}
\par~\par\noindent\centerline{e-mail:
Andrea.Tartari@mib.infn.it }

   \section*{Abstract} Our three frequencies radiometer MASTER, which allows
   low noise observations in three frequency intervals around 90,
   220 and 345 GHz is being completed. We discuss the possibility
   of exploiting the MASTER's characteristics for studies of the
   Sunyaev Zel'dovich effect from the Antartic Plateau and propose
   an observational program from Dome C.

\section{Introduction}

The Sunyaev-Zel'dovich (SZ) effect has been recognized, since the
time of its discovery, as a fundamental probe of the physical
processes in the \emph{intracluster medium} (ICM), and more and
more as a key instrument for the understanding of the evolution of
the Universe. It's well known that the SZ effect is also a source
of CMB secondary anisotropies and, with respect to this kind of
cosmological observations, SZ wide surveys are planned in near
future both from space and from ground observatories.\noindent

We propose a multifrequency observation based on the employment of
three heterodyne SIS receivers, with the aim of characterizing the
spectral signature of the thermal SZ. A detailed knowledge of the
spectrum, in turn, provides us two different ways to measure the
temperature of the CMB at cluster's redshift and it's necessary to
obtain a measure of the pure kinematic SZ.

\section{Astrophysics around 220 GHz}

\subsection{The thermal SZ}
When CMB photons interact with the thermal electrons of the
intracluster medium \emph{via} inverse Compton scattering, a
spectral distortion arises in the Rayleigh Jeans and in the Wien
region of the 3K black-body spectrum: the brightness temperature
in the RJ region decreases, while the Wien tail is characterized
by an enhancement. Between these spectral regions there's a
peculiar frequency (the crossover frequency) in which this effect
vanishes and, introducing the non dimensional frequency $$x=(h\nu
/k_BT_{CMB}),$$ this frequency gets the value $x_{0nr} \cong
3.830$ (see Fig.1 left panel).\noindent

Statistically, only the 1\% of the photons undergo Compton scattering, and
this results in a typical spectral distortion of the CMB thermodynamic
temperature on the order of $\frac{\Delta T}{T}\sim 10^{-4}$. This effect,
derived for the first time by Sunyaev and Zel'dovich \cite{zeldovich}
starting from the non relativistic Kompaneets equation, is known as non
relativistic thermal SZ effect (thSZ). Analytical descriptions of the
relativistic corrections to this effect have been developed in the last
decade \cite{Itoh} and, as pointed out by Rephaeli \cite{rephaeli a}, these
corrections are no longer negligible when $T_e>5 keV $. In particular, they
can affect significantly the spectral signature of the thermal SZ effect
shifting $x_{0nr}$ \cite{rephaeli b} and reducing the amplitude of the
spectral distortion. The first order corrected expression for $x_0$ is

\begin{equation}\label{1}
  x_0 \cong 3.83\big [1+1.1674\big (\frac{k_BT_e}{m_ec^2}\big )\big].
\end{equation} \noindent

More recently, a deep interest has risen on the non thermal SZ
(\cite{blasi}), which in principle could give us a new insight on the
relativistic electrons in the extended radio halos of clusters of galaxies.
Again, this effect should give a contribution to the SZ effect near $x_0$
(affecting $x_0$ itself), but it's extremely weak (\cite{shimon}), and its
detection seems unlikely in a near future.

\subsection{The peculiar motion of clusters}
The kinematic SZ (kSZ) is the intensity change due to the peculiar
motion of the cluster along the line of sight in the CMB rest
frame. The change in temperature is frequency independent, and it
is expressed as
\begin{equation}\label{2}
 \Delta T_{kin} = -T_{CMB}\frac{V_p}{c}\tau
\end{equation}
where $\tau$ is the optical depth of the electron plasma and $V_p$
the peculiar velocity, the sign depending on the velocity
direction. \noindent

The channel where the probability to measure this effect maximizes is the one
centered around $x_0$, simply because the thermal one vanishes there. This
means that the residual SZ at the crossover is purely kinematic, allowing us
to recover the peculiar velocity of the cluster once we know $\tau$
\cite{sunyaev}.\noindent

Nevertheless this effect is very weak and, from the best fit parameters
($T_e$,$\tau$ and $V_p$) derived for Abell 2163 by Hansen \cite{hansen}, we
deduce a temperature shift of the order of magnitude of $10\mu K$ (see Fig.1
right panel).\noindent

\begin{figure*}
   \centering
   \resizebox{\hsize}{!}{\includegraphics[clip=true]{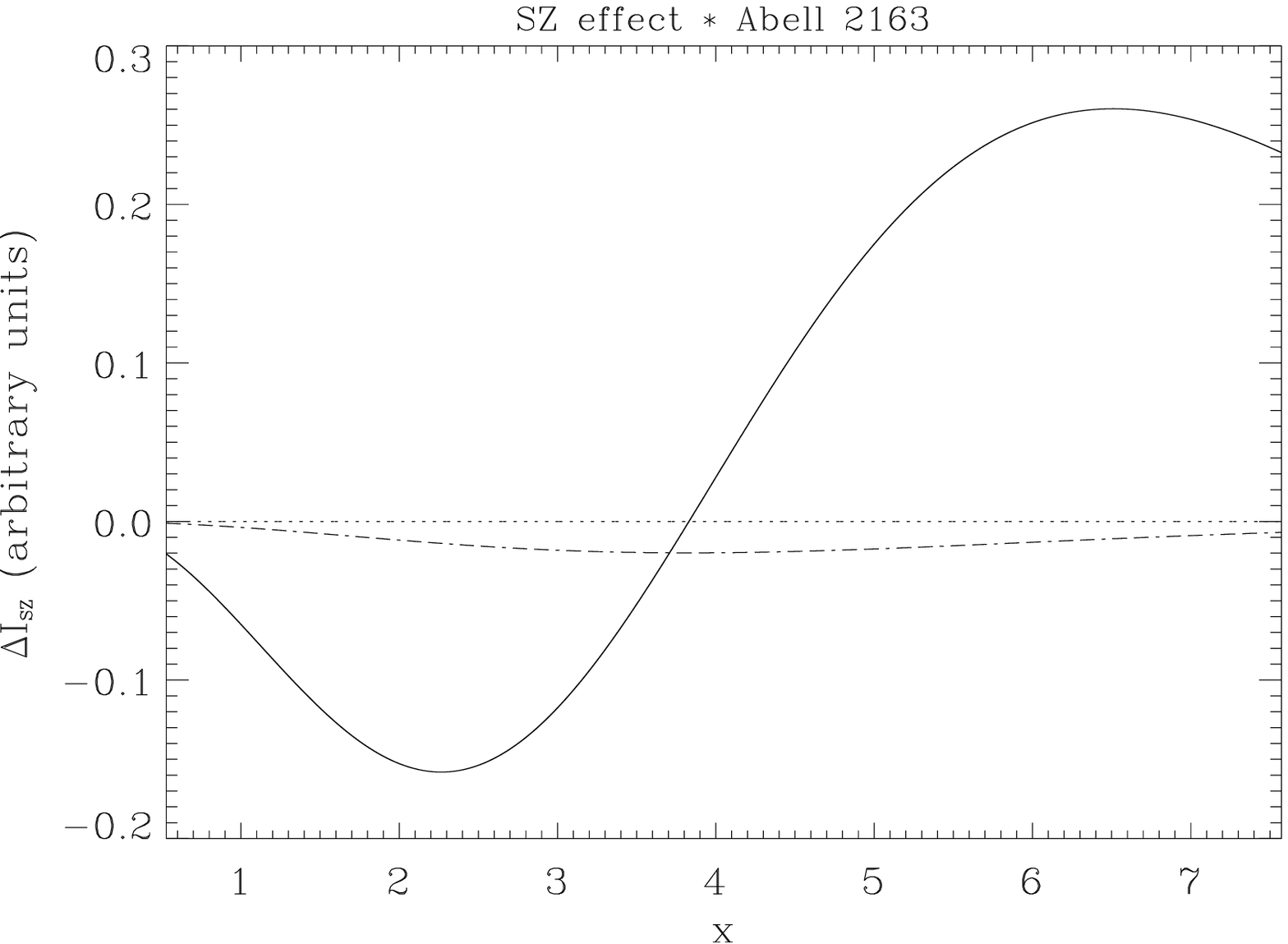}
   \includegraphics[clip=true]{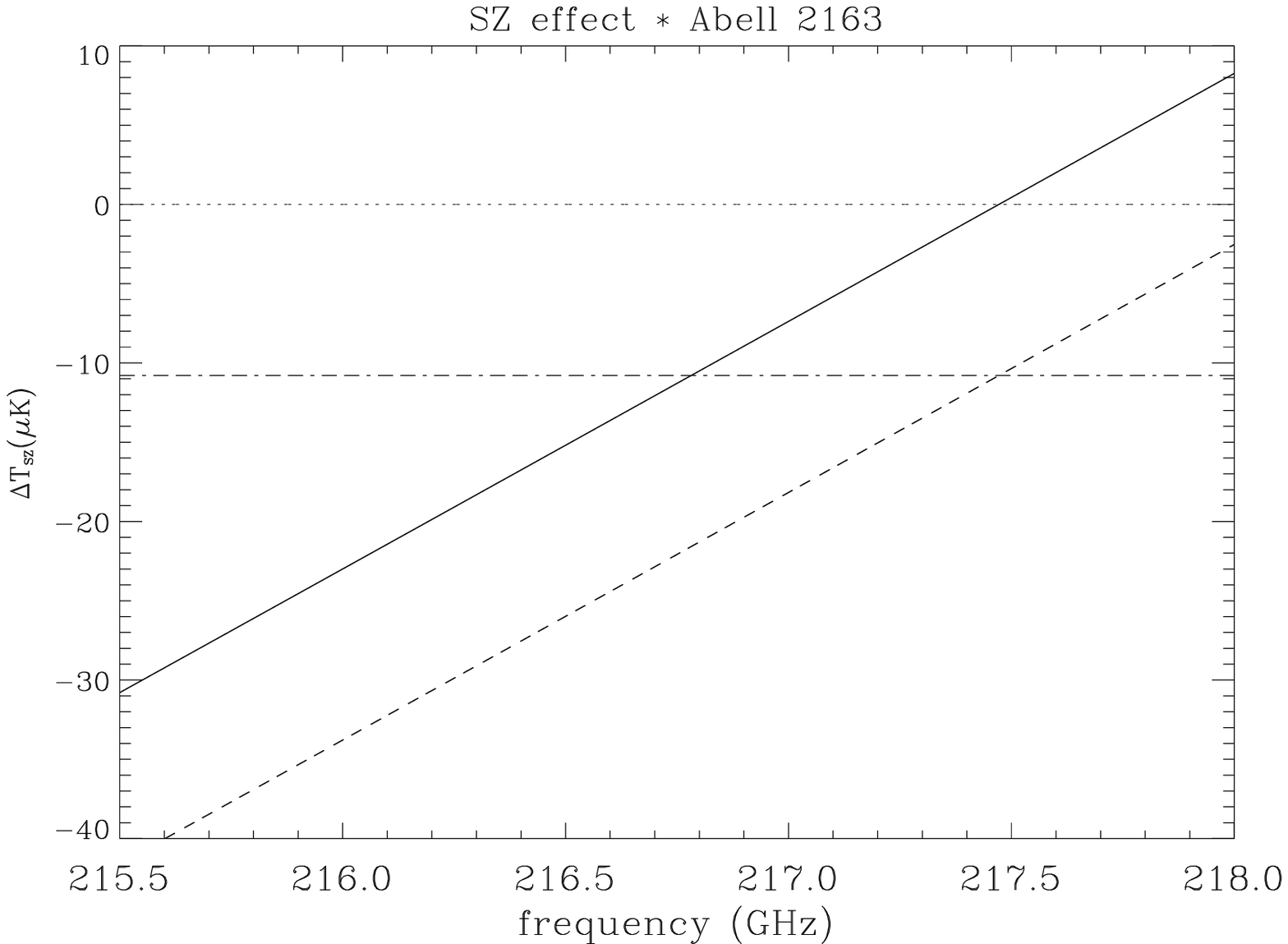}}
     \caption{Left panel: The non relativistic thermal (solid line) and kinematical (dash dot line) SZ effects deduced from best fit parameters
      derived by Hansen \cite{hansen} for Abell 2163. This is the spectral shape of intensity fluctuation in arbitrary
      units, with the kSZ enhanced by a factor 10. Right
      panel: The temperature change ($\mu K$) due to SZ is plotted \emph{versus} frequency
      ($GHz$) both for thermal and kinematical SZ effects (respectively solid and dash dot line). The dashed
      line represent the sum of the two effects. This plot is centered around $ \nu_{0nr}$.}
        \label{fit1}
\end{figure*}

Such a weak signal requires a high sensitivity in order to be
detected, and we know that the more sensitive receivers nowadays
are bolometric arrays cooled down to $300mK$ or less.
Nevertheless, if using wide band devices is recommendable in order
to reduce the integration time, it's also true that we need
frequency resolution around the crossover to disentangle the
kinematic from the thermal effect. Another reason which convinced
us to pay in sensitivity, with the aim of gaining in spectral
accuracy, is the necessity of reconstructing the global spectral
signature of thSZ.

\subsection{In situ measurment of $T_{CMB}(z)$}

A task strictly connected to the determination of the crossover frequency is
the measurment of the CMB temperature at cluster's redshift. The key idea is
that the non dimensional frequency, in a FLRW Universe, is redshift
independent, because $\nu$ and $T_{CMB}$ exibit the same $(1+z)$-scaling.
Since $x_0$ is fixed and supposing that we are able to find ${\nu}_0$, we can
go back to $T_{CMB}(z)$(\cite{fabbri}), providing a further test to the
standard cosmology (\cite{losecco}). Besides, we can exploit the increasing
quality of the X-ray cluster surveys (see the observation of the Coma cluster
with XMM by Arnaud \cite{arnaud}) in order to consider possible relativistic
corrections to $x_{0nr}\cong 3.830$, which require a detailed knowledge of
$T_e$. It's important to underline that, neglecting the relativistic
correction to $x_{0nr}$ for a hot plasma with $T_e\cong 15 keV$, the CMB
temperature derived in this way will be overestimated by few percent ($\sim
3\%$), since, according to eq.(1), there's a shift in frequency of $\sim 6$
GHz. A multifrequency approach to the measurment of $T_{CMB}(z)$, similar to
that proposed by Rephaeli \cite{rephaeli c} is also a viable chance. This
technique relies on the measurement of the thSZ in three frequency bands,
which allows one to build two ratios, $\Delta I_{\nu 1}/\Delta I_{\nu 2}$ and
$\Delta I_{\nu 3 }/\Delta I_{\nu 1}$, completely characterized in the
$(y,T_{CMB})$ parameter space. Recently Battistelli (\cite{battistelli})
reported a $T_{CMB}$ measurement in the Coma cluster and in Abell 2163 with
the MITO telescope, using this procedure.

\subsection{Foregrounds}
The main foreground in our frequency bands is dust emission both
from Galactic Cirrus and, eventually, from IC dust, since we can
not rule it out \emph{a priori}.

The presence of dust in galaxy clusters seems to be likely because of the
galactic winds responsible of the enrichment of the ICM \cite{faber}. In this
context, the survival of dust grains of galactic origin under sputtering in a
hot plasma has been extensively discussed (as reference,
\cite{burke}).\noindent

In order to achieve an order of magnitude of the brightness temperature
($T_{Bd}$) of dust grains in clusters, we followed the model of dust emission
built by Dwek \cite{dwek} for the Coma cluster. This model predict a $100\mu
m$ brightness $\approx 0.2 MJy sr^{-1}$ for the cluster center (the region
inside $R<2Mpc$) where $T_d\cong 20 K$, $T_d$ being the dust physical
temperature. Normalizing a modified black-body spectrum with respect to this
value and extrapolating to lower frequencies, we found that this foreground
emission is comparable to the kinematic SZ effect around $220 GHz$ ($\sim 3
\mu K$). On the other hand, we adopted a frequency dependent emissivity
$\varepsilon (\nu)\sim \nu^{\alpha}$ with $\alpha=1.5$, even if a steeper
spectrum ($\alpha \simeq 2.0$) seems preferable. This means that we obtained
an upper limit for $T_{Bd}$ (since we overestimated the brightness in the low
frequency tail), and that we can guess that this foreground won't affect thSZ
measurements, even if it could affect significantly kSZ.\noindent

The far infrared (FIR) emission of the Galactic Cirrus can be in principle
the dominant foreground, even if the observational techniques help us to keep
its impact under control, as described in \ref{The observations}. Besides,
the increasing quality of CMB data is providing us with a detailed
description of galactic foregrounds (\cite{masi}, \cite{fink}), whose
anisotropic components cannot be removed with differential sky-chopping
techniques. On the other hand, we could exploit our three-frequency
measurements to disentangle the dust emission from the cosmological signal.

\section{Observations with MASTER}
\subsection{The receiver}
MASTER is a triple heterodyne receiver operating at three central
frequencies corresponding to atmospheric windows: $94$, $220$,
$345$ GHz. \noindent

The front-end of the system, cooled to 4 K, is characterized by
three SIS mixers, which are the only devices allowing coherent
detection above $200$ GHz, followed by low noise HEMT amplifiers
operating in the range $1\div 10$ GHz, the IF filters being
centered at $1.2$ GHz with $\pm 1$ GHz bandwidth. The details
concerning the receiver and the optical coupling with the MITO
telescope at the Testagrigia observatory have been discussed in
previous works (\cite{battistelli2} and references therein). Now
it is useful to recall that the expected noise temperatures of the
SIS at 94, 220 and 345 GHz are respectively $\sim 120$, $\sim 140$
and $\sim 160$ K. The tuning bandwidhts are $10 \% $ of the
central frequencies, while the instantaneous bandwidth is $0.6$
GHz.

\subsection{The observations}
\label{The observations}

Our aim is to study galaxy clusters, structures which extend in
the sky with diameters ranging from few arcminutes to about ten
arcminutes. This means that we can achive the suitable resolution
for this kind of observations with a $2$-m class telescope (like
MITO or OASI). The optimal observing technique for SZ measurments
is a three-field modulation of the sky signal, obtained by
chopping the secondary mirror of the telescope. In this way, we
observe alternatively the cluster and two reference patches of sky
diametrally opposed with respect to the cluster itself (ON-OFF
modulation followed by synchronous detection). Moreover, the beam
switching is carried out at constant elevation, in order to cut
(on average) the atmospheric emission. It is also important to
remind that the angular amplitude of the modulation has to be
chosen in such a way to guarantee that the beam in the
OFF-position shall be completely outside the cluster. This fact,
in turn, sets a minimum value for the amplitude $\Delta \theta$ of
sky chopping.
\begin{figure*}
   \centering
   \resizebox{\hsize}{!}{\includegraphics[clip=true]{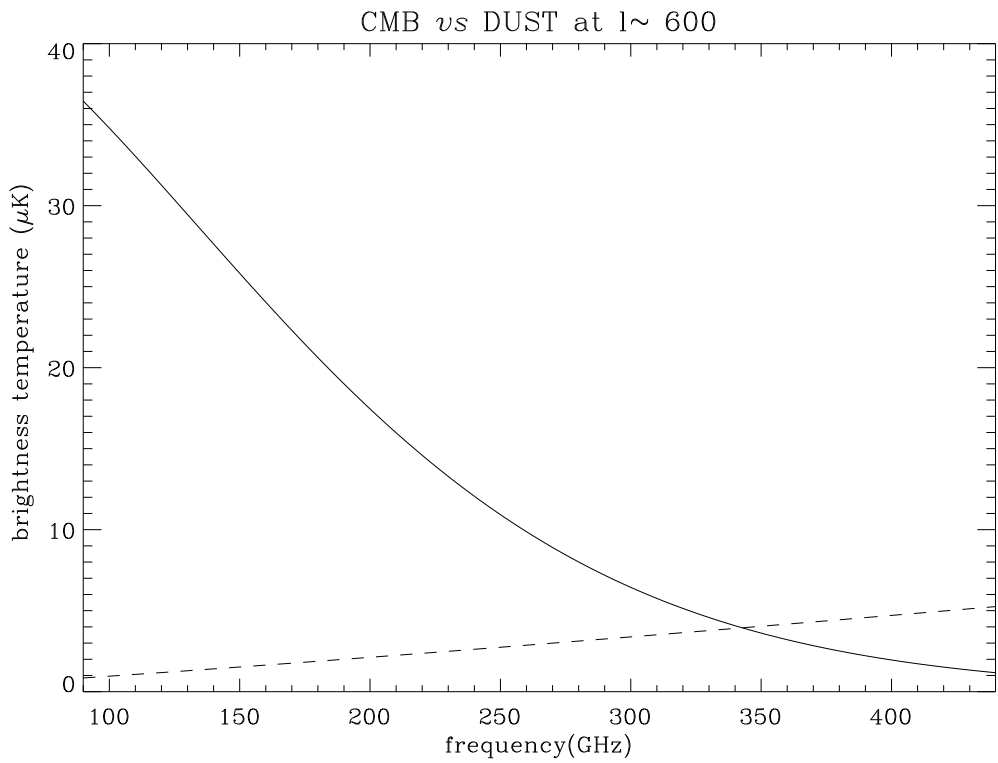}}
     \caption{The brightness temperature of CMB in $\mu K$ (solid line) compared with the dust
     anisotropic emission at medium galactic latitudes (dashed line) at the multipole $l\simeq 600$.}
        \label{fit}
\end{figure*}
This differential technique has the advantage that the foregrounds with
multipole index $0\leq l<(\pi/\Delta \theta)$ are removed, and that gain
variations could be kept under control. On the other hand, the sensitivity of
the instrument is reduced, because half of the observing time is spent on the
reference regions in the sky. It's important to underline that the
anisotropic component of the galactic dust emission and the CMB anisotropies
(on the angular scale corresponding to $\Delta \theta$) are not cut by the
sky signal modulation and, statistically, they will affect SZ detection. In
particular, different chopping amplitudes will be associated to different
correlated dust signals, because the interstellar dust exhibits an angular
power spectrum scaling as a power law $c_l\sim l^{-\beta}$ with $2 \leq \beta
\leq 3$, depending on the galactic latitude (\cite{masi}). At the same time,
our experiment will be liable to the multipole $l\sim 1/\Delta\theta$ of CMB
anisotropies. We noted that a beam switching of $\sim 20'$ amplitude
introduces an overall $\sim 20\mu K$ noise level around $220$ GHz, mainly due
to the CMB (see fig.2). Again, this effect overtakes the expected kSZ signal.

\section{Conclusions}

We propose two main targets to be achieved with the MASTER
receiver: the \emph{in situ} measurement of the CMB thermodynamic
temperature and the detection of the kinematic SZ effect. Both
these measurements are linked to the spectral characterization of
thSZ in three frequency bands, which could be achieved with a
remarkable accuracy exploiting the narrow instantaneous bandwidth
of our receiver and the chance of LO tuning (particularly around
the expected crossover frequency). Nevertheless, passing from the
first task to the second one, a jump of more than a factor $10$ in
sensitivity is required. We know also that this jump cannot be
achieved simply extending the integration time, because long-term
instabilities could arise in the electronics.\noindent

The electronic instabilities become decisive in limiting the sensitivity when
the spectral density of the $1/f^{\alpha}$ noise equals the pure white noise
contribution \cite{wollack}. This happens in correspondence of a
characteristic frequency, $f_{knee}$, which depends on the intrinsic
properties of the device. The knowledge of the knee frequency is of great
practical importance, because it is used to set a lower limit to the chopping
frequency. In turn, this value has to be compared with the lower limit set by
atmospheric instabilities. In fact, we know that long time integration is
also exposed to sky-noise. A first condition to be accomplished to face the
latter problem consists in carrying out the observations in a site with a low
PWV content and stable atmospheric conditions (with regard to this, Dome C
seems to be the ideal site, \cite{valenziano}). Then, the chopping period
should be lower than the typical timescale of atmospheric fluctuations, and
the beam switching angle should be smaller than the angular scale of
correlated atmospheric emission.\noindent

As final remark, particularly thinking to an observational
campaign from Dome C, we add that the choice of the telescope's
optics will allow us to estimate a realistic value for $T_{sys}$
(and hence for the sensitivity) and it will suggest us also the
final optical configuration of MASTER.

\bibliographystyle{aa}

\end{document}